%% file: main.tex
\documentclass[9pt, conference]{IEEEtran}
\IEEEoverridecommandlockouts
\usepackage{cite}
\usepackage{amsmath,amssymb,amsfonts}
\usepackage{algorithmic}
\usepackage{graphicx}
\usepackage{textcomp}
\usepackage{xcolor}
\usepackage{algorithm}
\usepackage{multirow}
\usepackage{bm}
\usepackage{booktabs}
\renewcommand{\baselinestretch}{0.893}

\def\BibTeX{{\rm B\kern-.05em{\sc i\kern-.025em b}\kern-.08em
    T\kern-.1667em\lower.7ex\hbox{E}\kern-.125emX}}
\begin{document}

\title{An Efficient Asynchronous Batch Bayesian Optimization Approach for Analog Circuit Synthesis}
\author{\normalsize{Shuhan Zhang$^1$, Fan Yang$^{1*}$, Dian Zhou$^2$ and Xuan Zeng$^{1*}$} \\
\normalsize{$^1$State Key Lab of ASIC \& System, School of Microelectronics, Fudan University, Shanghai, P. R. China} \\
\normalsize{$^2$Department of Electrical Engineering, University of Texas at Dallas, Richardson, TX, U.S.A.}\vspace{-3.5mm}\\
\thanks{*Corresponding authors: \{yangfan, xzeng\}@fudan.edu.cn.}}

\maketitle

\input{include/abstract}
\input{include/introduction}

\input{include/background}
\input{include/proposal}

\input{include/experiment}
\input{include/conclusion}

\section*{Acknowledgements}
This research is supported partly by the National Major Science and Technology Special Project of China (2017ZX01028101-003), partly by National Natural Science Foundation of China (NSFC) 61822402, 61774045, and 61929102.

\bibliographystyle{unsrt}
\bibliography{bibliography}

\end{document}

%% file: include/abstract.tex
\begin{abstract}
In this paper, we propose EasyBO, an Efficient ASYnchronous Batch Bayesian Optimization approach for analog circuit synthesis. In this proposed approach, instead of waiting for the slowest simulations in the batch to finish, we accelerate the optimization procedure by asynchronously issuing the next query points whenever there is an idle worker. We introduce a new acquisition function which can better explore the design space for asynchronous batch Bayesian optimization. A new strategy is proposed to better balance the exploration and exploitation and guarantee the diversity of the query points. And a penalization scheme is proposed to further avoid redundant queries during the asynchronous batch optimization. The efficiency of optimization can thus be further improved. Compared with the state-of-the-art batch Bayesian optimization algorithm, EasyBO achieves up to 7.35$\times$ speed-up without sacrificing the optimization results.
\end{abstract}

%% file: include/introduction.tex
\section{Introduction}

As device size scales to the nano region, it becomes difficult to manually design analog circuits. With the increasing variation space and the narrowing time-to-market, a sophisticated analog circuit optimization algorithm is in great need to improve the productivity. Typically, the analog circuit design procedure can be divided into two parts: topology selection and device sizing \cite{rutenbar2007hierarchical}. In this paper, we focus on the device sizing problem with a fixed circuit topology.

To meet the design specifications in a short period of time, the overall simulation time which dominates the cost of the optimization process should be reduced. Traditional optimization algorithms for device sizing fall into two categories: the model-based and the simulation-based methods. The model-based method investigates the circuit performance with analytical equations or polynomial models. By approximating the behavior of the analog circuit, the constructed model substitutes the computationally expensive simulator to search the global optimum. One well-known model-based approach is geometric programming \cite{kim2004techniques, mandal2001cmos}, which models the circuit performance with posynomial approximation \cite{li2004robust, daems2002efficient, eeckelaert2003generalized}. However, the optimization results of the model-based method largely depend on the modeling accuracy. The performance deviation between the constructed model and the real circuit would make the obtained results diverge from the real optimum. 

The simulation-based approaches instead treat the behavior of the analog circuit as a black-box function. Based on the current simulated dataset, they explore the state space by proposing the next locations for evaluation. There are many well-developed simulation-based algorithms, including the multi-starting point (MSP) algorithm \cite{yang2018smart-msp:, peng2016efficient, lv2017subgradient}, simulated annealing (SA) \cite{gielen1990analog, phelps2000anaconda:, grzechca2007use}, differential evolution (DE) \cite{liu2009analog}, and particle swarm optimization (PSO) \cite{fakhfakh2010analog, wu2009a, vural2012analog, coello2003use}. However, a serious defect that hinders simulation-based approaches from being widely used is its relatively low convergence rate.

To solve this dilemma, the Bayesian optimization algorithm (BO) has been recently proposed to accelerate the optimization procedure by combining both the model-based and simulation-based strategies \cite{lyu2017efficient}. 
Instead of constructing the model once and search the design space offline, BO refines the surrogate model incrementally and invokes the simulation on the fly. Despite the debuts of recent works \cite{zhang2019efficient, zhang2019bayesian, lyu2017efficient, lyu2018multi} that have proven the effectiveness and efficiency of the BO framework, the sequential characteristic of the state-of-the-art acquisition function in the BO framework makes it hard to be parallelized. 
Without parallelism, the hardware is not able to be fully utilized when multi-core workstations are available. 
Also, for circuits that are expensive to evaluate, it is much more cost-effective to simulate the circuits in a batch rather than one-at-a-time. 
In order to further reduce the overall time consumption on the simulation, efforts have been made to make batch BO possible \cite{lyu2018batch, hu2018parallelizable}. 
By synchronously sampling several points at each iteration and evaluating the performances in parallel, they are able to greatly reduce the overall simulation time compared to the sequential BO.

Nevertheless, the main issue with the synchronous batch BO is that different design parameters can lead to different simulation time consumption. 
And it is a waste of hardware resources to let workers wait idly for the slowest design in the batch to finish the simulation. 
To fill this gap, we propose EasyBO, an asynchronous batch BO algorithm for analog circuit synthesis. 
Instead of waiting for the whole batch to finish the evaluation process, EasyBO issues a new candidate point whenever there is a worker becomes available in the current batch.
We introduce a new acquisition function which can better explore the design space for asynchronous batch Bayesian optimization. EasyBO introduces a new strategy to better balance the exploration and exploitation and guarantees the diversity of the query points during the asynchronous batch Bayesian optimization. The design space can thus be better traversed and thus the optimization efficiency can be greatly improved. 
%
A penalization scheme is proposed to further avoid redundant queries during the asynchronous batch optimization. 
%
EasyBO is experimentally evaluated on two real-world analog circuits.
Compared to DE method \cite{liu2009analog}, EasyBO can achieve up to 1935$\times$ speed-up while achieving better optimization results.
Compared with the state-of-the-art synchronous batch BO algorithm, experimental results demonstrate that EasyBO can achieve up to 7.35$\times$ speed-up with comparable optimization results. 

The rest of this paper is organized as follows. A brief review of the Gaussian process regression model based BO framework and the synchronous batch BO approaches is presented in \S\ref{sec:background}. Our proposed asynchronous batch BO algorithm are presented in \S\ref{sec:proposal}. The experimental results of two real-world analog circuits are compared with several state-of-the-art optimization algorithms in \S\ref{sec:experiment}. And we conclude the paper in \S\ref{sec:conclusion}.

%% file: include/background.tex
\vspace{-0.1cm}
\section{background} \label{sec:background}

In this section, we first present the problem formulation of the analog circuit optimization (\S\ref{sec:problem_formulation}). We then give a brief review of both the Gaussian process regression model based BO framework (\S\ref{sec:BO}) and the synchronous batch BO (\S\ref{sec:sync_BO}).

\vspace{-0.1cm}
\subsection{Problem Formulation} \label{sec:problem_formulation}

For an analog circuit, there are typically several circuit performance metrics to be optimized simultaneously. In this paper, we formulate the analog circuit optimization problem into a single-objective optimization problem by assigning each performance metric a weight:
\begin{equation}
    \text{maximize.} \quad \text{FOM}(\bm{x}) = \sum_{i=1}^{m} \alpha_i f_i(\bm{x}), 
\end{equation}
where $\bm{x}$ is a $d$-dimensional design variables, FOM$(\cdot)$ represents the Figure of Merit for optimization, $f_i(\cdot)$ denotes the $i$-th performance metric and $\alpha_i$ is the corresponding assigned weight. Our proposed approach can also be easily extended to handle constrained optimization problem, which will be discussed in future work. 

\vspace{-0.1cm}
\subsection{Bayesian Optimization} \label{sec:BO}

Generally, there are two basic elements in the Bayesian optimization (BO) framework: the surrogate model and the acquisition function \cite{rasmussen2003gaussian}. 
The surrogate model captures our prior belief about the unknown objective function and provides the posterior distribution with both predictive mean and uncertainty estimations \cite{shahriari2015taking}. 

One most frequently used surrogate model is the Gaussian process regression (GPR) model \cite{rasmussen2003gaussian}. For a given scalar-valued latent function $f(\cdot)$ defined over a compact space $\chi: \mathbb{R}^d \rightarrow \mathbb{R}$, we assume the observation noise $\epsilon \sim N(0, \sigma^2_n)$.
By assuming $f \sim \mathcal{GP}(m(\bm{x}), k(\bm{x}, \bm{x}))$, we encode our prior belief about the unknown objective function with a mean function $m(\cdot)$ and a kernel function $k(\cdot, \cdot)$. In this paper, we set the kernel function as the square exponential function: $k_{SE}(\bm{x}_i, \bm{x}_j) = \sigma^2_f exp(-\frac{1}{2}(\bm{x}_i - \bm{x}_j)^T \Lambda^{-1} (\bm{x}_i - \bm{x}_j))$, where $\Lambda=diag(l_1^2, \dots, l_d^2)$ is a diagonal matrix, $l_i$ is the length scale in $i$-th dimension, and $\sigma_f$ represents the variance. 

Let $D=\{X, \bm{y}\}$ denotes a dataset with N points, $X=\{\bm{x}_1, \dots, \bm{x}_N\}$ is the input locations and $\bm{y}=\{y_1, \dots, y_N\}$ is the corresponding observations. The posterior distribution for a given input design $\bm{x}_\ast$ is \cite{rasmussen2003gaussian}:
\begin{equation}
\label{eq:prediction}
    \begin{cases}
	& \mu(\bm{x}_\ast) = k(\bm{x}_\ast, X) K^{-1} \bm{y}\\
	& \sigma^2(\bm{x}_\ast) = k(\bm{x}_\ast, \bm{x}_\ast) - k(\bm{x}_\ast, X) K^{-1} k(X, \bm{x}_\ast), 
    \end{cases}
\end{equation}
where $k(\bm{x}_\ast, X)=\{k(\bm{x}_\ast, \bm{x}_1), \dots, k(\bm{x}_\ast, \bm{x}_N)\}$, $k(X, \bm{x}_\ast) = k(\bm{x}_\ast, X)^T$, and $K=k(X, X)+\sigma^2_nI$ is the covariance matrix.

The acquisition function works as a utility function that provides the fitness value for each candidate point, and selects the query point to refine our prior belief \cite{shahriari2015taking}. By carefully balancing between the exploration and exploitation, the acquisition function efficiently search the design space by avoiding two pitfalls: (1) too much exploration in the low potential area, (2) too much exploitation around the sampled region. One commonly used acquisition function is the upper confidence bound (UCB) \cite{srinivas2009gaussian}:
\begin{equation} \label{eq:UCB}
    UCB(\bm{x}) = \mu(\bm{x}) + \kappa * \sigma(\bm{x}),
\end{equation}
where $\kappa$ is the parameter characterizing the trade-off between the exploration and exploitation. There are also many other well-developed acquisition functions, including expected improvement (EI) \cite{mockus1978application}, probability of improvement (PI) \cite{kushner1964new}, entropy search (ES) \cite{hennig2012entropy} and Thompson sampling (TS) \cite{thompson1933likelihood}. Extensive researches have proven that a portfolio of several acquisition functions is also possible \cite{hoffman2011portfolio}. 

In summary, BO constructs the initial surrogate model with a set of randomly sampled dataset. By leveraging the predictive mean and uncertainty estimation, the acquisition function selects the next query point with the maximum fitness value to incrementally refine the surrogate model. After a certain number of iterations, the global optimum will be reached with a theoretical guarantee.

\vspace{-0.1cm}
\subsection{Synchronous Batch Bayesian Optimization} \label{sec:sync_BO}

In order to improve the utilization of the hardware resources, many different synchronous batch BO approaches have been proposed to make evaluation in parallel possible, including MACE \cite{lyu2018batch}, pBO \cite{hu2018parallelizable}, pHCBO \cite{hu2018parallelizable}, BUCB \cite{desautels2014parallelizing}, and LP \cite{gonzalez2016batch}. 
Although parallelism can help gather more information in a short period of time, it also introduces the information gap when selecting future locations without intermediate observations. Most of the synchronous batch BO algorithms convert the batch selection into a sequential procedure, since it is hard or even computationally intractable to maximize the sample efficiency while maintaining diversity. MACE \cite{lyu2018batch} maintains diversity for each batch by sampling from the Pareto front of the multi-objective acquisition function ensemble. LP \cite{gonzalez2016batch} penalizes the acquisition function in the neighborhood of the selected locations. BUCB \cite{desautels2014parallelizing} penalizes around the busy locations by using hallucinated observations. pBO \cite{hu2018parallelizable} tries to introduce diversity to the batch selection by assigning the predictive mean and uncertainty measurement with different weighting parameters, which can be expressed as \cite{hu2018parallelizable}:
\begin{equation} \label{eq:pbo}
    \alpha_{pBO}(\bm{x}, w_i) = (1-w_i)*\mu(\bm{x})+w_i*\sigma(\bm{x}), 
\end{equation} 
where $\bm{w}=(0, 0.25, 0.5, 0.75, 1)^T$ when batch size is 5. Higher $w_i$ means that we tend to visit the regions with higher uncertainties (exploration). Lower $w_i$ means that we tend to traverse better solutions with high confidence (exploitation). Thus, different $w_i$ corresponds to different trade-offs between exploration and exploitation. The basic idea of \cite{hu2018parallelizable} is to fully explore different trade-offs with a set of different $w_i$ in parallel. However, the query points of explicitly designed $w_i$ in \cite{hu2018parallelizable} still have high probability to fall into the same region, which would greatly reduce the efficiency of exploring the state space. 

In order to address this problem, pHCBO \cite{hu2018parallelizable} penalizes the acquisition function of pBO with an additional term $\alpha_{HC}(\bm{x}, D_b, w_i)$
\begin{equation}
    \alpha_{pHCBO}(\bm{x}, w_i) = \alpha_{pBO}(\bm{x}, w_i) - \alpha_{HC}(\bm{x}, D_b, w_i), 
\end{equation} 
where
\begin{equation} \label{penalize}
    \alpha_{HC}(\bm{x}, D_b, w_i) = N_{HC} \sqrt[5]{\prod^5_{j=1} exp[(\frac{d}{d_{\bm{x}}})^{10}]}.
\end{equation}

Here $D_b$ denotes the observed dataset for $b$-th iteration, $d$ is a manually defined parameter, $d_{\bm{x}}=\| \bm{x}-\bm{x}_{b-j,i}\|$, and $\bm{x}_{b-j,i}$ is the query location at $(b-j)$-th iteration for $w_i$. If $\bm{x}$ falls into the neighborhood of the previous 5 points, the penalization term would be extremely large. The penalization term is thus possible to prevent clustered sampling by the same acquisition function (with the same $w_i$ value).

%% file: include/proposal.tex
\vspace{-0.1cm}
\section{Proposed Approach} \label{sec:proposal}

In this section, we will present the motivation and challenges of asynchronous batch Bayesian optimization (\S\ref{sec:motivation}) firstly. Then, we will describe our carefully designed acquisition function (\S\ref{sec:acq}) and penalization scheme for asynchronous batch Bayesian optimization (\S\ref{sec:penalization}). Finally, we summarize EasyBO algorithm (\S\ref{sec:summary}). The batch size is denoted as $B$ throughout the rest of the paper.

\vspace{-0.1cm}
\subsection{Asynchronous Batch Bayesian Optimization} \label{sec:motivation}

\begin{figure}
    \centering
    \includegraphics[width=0.35\textwidth]{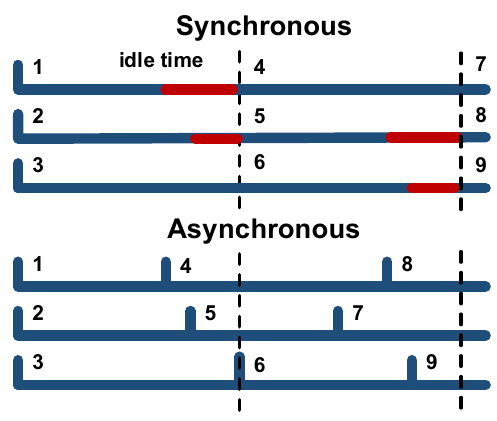}
    \vspace{-0.4cm}
    \caption{An illustration of asynchronous and synchronous setting when batch size is 3.}
    \label{fig:asynchronous}
    \vspace{-0.5cm}
\end{figure}

As illustrated in Figure \ref{fig:asynchronous}, the synchronous batch BO algorithm aims to select a batch of candidate points to evaluate in parallel. The next batch will only be issued when all samples in the previous batch have been evaluated. Due to variability in simulation times for different design parameters, there will always be workers waiting idly for others to finish their jobs in the synchronous batch BO algorithm. Therefore, the synchronous batch BO algorithm is not able to achieve $B\times$ speed-up for batch size $B$ compared to its sequential counterpart. As the batch size increases, the time reduction effect will deteriorate quickly, since more workers will wait idly for others to finish their jobs. 



Instead of waiting for the whole batch to finish the evaluation process, we propose to asynchronously issue new candidate points whenever a worker becomes available. The key motivation behind the asynchronous batch BO algorithm is to make full use of the hardware resources to reduce the overall time spent on evaluation. Intuitively, the asynchronous batch BO algorithm can process a greater number of evaluations than its synchronous counterpart, in a given period of time. With the increase of the batch size, the time reduction for a given number of simulations will be more significant compared to its synchronous counterpart. And the asynchronous batch BO algorithm is especially suitable for problems where the simulation time of which differs greatly for different design parameters. To the best of our knowledge, this is the first time when an asynchronous mechanism is proposed for Bayesian optimization of analog circuits.


In general, there are two harsh challenges for the batch BO algorithms to deal with: (1) how to fully leverage our current knowledge about the latent function, and select the future query points, (2) how to penalize around the selected locations that are still under evaluation to prevent redundant samples from being chosen in the busy region.

\begin{figure}
    \centering
    \includegraphics[width=0.35\textwidth]{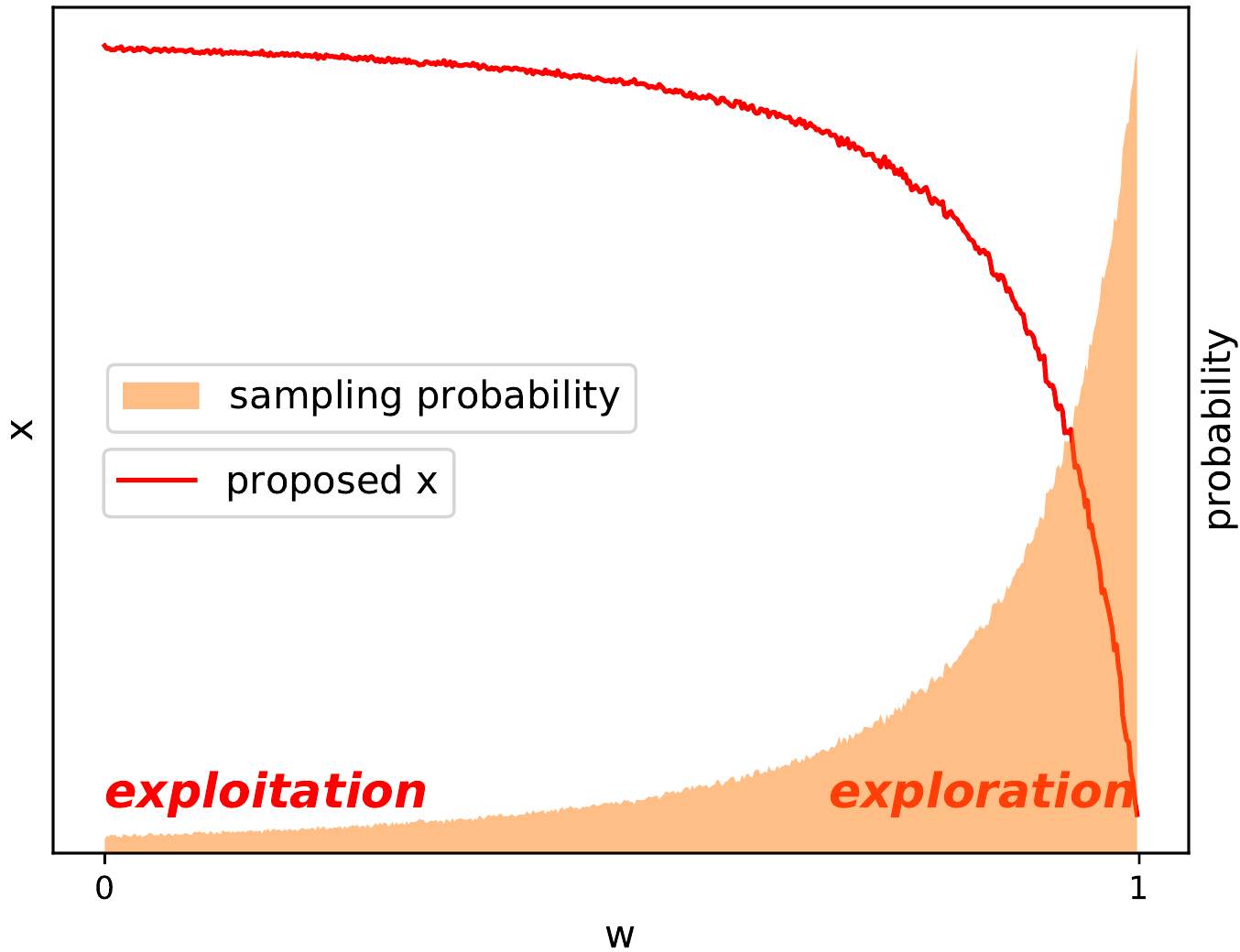}
    \vspace{-0.4cm}
    \caption{An illustration of the UCB acquisition function with different $w$ over [0,1]. And the sampling probability of our proposed acquisition function with respect to $w$.}
    \label{fig:acquisition}
    \vspace{-0.5cm}
\end{figure}

\vspace{-0.1cm}
\subsection{Improved Acquisition Function} \label{sec:acq}
In Bayesian optimization, the acquisition function is carefully designed to balance the exploration and exploitation. For batch Bayesian optimization, it is also desirable for the acquisition function to create the diversity of the query points in a batch. The upper confidence bound (UCB) as shown in (\ref{eq:UCB}) is a direct yet powerful acquisition function. As explained in \S\ref{sec:sync_BO}, the predictive mean $\mu(\bm{x})$ represents the exploitation, while the predictive uncertainty $\sigma(\bm{x})$ represents the exploration. The parameter $\kappa$ is introduced to balance the exploration and exploitation.

In \cite{hu2018parallelizable}, the diversity is introduced to the batch selection by assigning the predictive mean and uncertainty measurement with different weighting parameters as shown in (\ref{eq:pbo}). We rewrite (\ref{eq:pbo}) here 
\begin{equation} \label{eq:pbo_re}
    \alpha_{pBO}(\bm{x}, w) = (1-w)*\mu(\bm{x})+w*\sigma(\bm{x}).
\end{equation} 
For batch size $B$, $B$ weights $\{w_1, \cdots, w_B\}$ which are uniformly distributed over $[0,1]$ are selected. With $B$ different weights, the $B$ different acquisition functions are expected to select $B$ different query points for a batch. However, such a strategy does not work well in real applications.

At the starting stage of the optimization procedure, it is important to gather global information about the underlying behavior of the objective function. In other words, the acquisition function should encourage exploration at the initial stage of the optimization process, which means $w$ should be larger to encourage exploration. 

After a limited number of iterations, the predictive uncertainty $\sigma(\bm{x})$ of the refined model would have a much smaller magnitude than its predictive mean $\mu(\bm{x})$. Therefore, the acquisition functions with smaller $w$ would generate almost the same query points, since $(1-w)*\mu(\bm{x})$ dominates the acquisition function in (\ref{eq:pbo_re}). From the previous analysis, we should encourage larger $w$ for the acquisition function as shown in (\ref{eq:pbo_re}). The uniformly distributed $w$ in~\cite{hu2018parallelizable} is not a good choice.

The corresponding distribution of the selected location $\bm{x}$ with respect to $w$ parameter is shown in Figure \ref{fig:acquisition}. It is worth noting that $\bm{x}$ only has small change when $w$ is relatively small and encourages exploitation. And the value of $\bm{x}$ changes quickly when $w$ is relatively large and encourages exploration. Therefore, we need to increase the sampling density around the region with large $w$ to maintain the diversity of the selected query points.

In EasyBO, instead of uniformly sampling the $w$ parameter, we propose a new acquisition function scheme that increases the sampling density as $w$ increases, which can be expressed as:
\begin{equation} 
    \alpha(\bm{x}, w) = (1-w)*\mu(\bm{x})+w*\sigma(\bm{x}),
\end{equation} 
where $w=\kappa/(\kappa+1)$. By randomly sampling $\kappa$ from the uniform distribution over a proper range $[0, \lambda]$, $w$ tends to approach 1 to encourage exploration at the initial stage and the diversity at the later optimization stage. The corresponding sampling probability of $w$ is shown in Figure \ref{fig:acquisition}. It can be observed that $w$ has higher probability near the neighborhood of 1. We set $\lambda$ to a limited value to prevent the acquisition function from too much exploration during the optimization. We set $\lambda=6.0$ in this paper.

\vspace{-0.1cm}
\subsection{Penalization Scheme} \label{sec:penalization}

In order to improve the optimization efficiency, penalization is proposed in \cite{hu2018parallelizable} to prevent the new query points from falling into the neighboorhood of the previous 5 points as shown in (\ref{penalize}). However, this penalization is not appropriate for batch Bayesian optimization. The purpose of the penalization for batch Bayesian optimization is to guarantee the diversity of the query points in one batch rather than the diversity of query points across the observation dataset. The penalization in (\ref{penalize}) would significantly impact the convergence rate. Note that the query points would concentrate in the neighboorhood of the optimal solution to guarantee the regression accuracy of GPR in the final stage of optimization. The penalization of diversity in (\ref{penalize}) would thus reduce the regression accuracy around the optimal solution and reduce the convergence rate. 

In this paper, we propose a new penalization scheme to guarantee the diversity of the query points of a batch. An interesting observation is that the predictive uncertainty of GPR model provides a natural penalization for diversity. The uncertainties are lower in the neighborhood of already sampled data points while larger in the unvisited region. Thus, the UCB acquisition function as shown in (\ref{eq:pbo_re}) can naturally avoid redundant sampling over the neighborhood of already visited points.

For batch Bayesian optimization, we should include the query points in the same batch to the training dataset, and incorporate the corresponding predictive uncertainty into the acquisition function to guarantee the diversity of the query points in one batch.

Denote $D=\{X, \bm{y}\}$ the observed data points. Denote $\hat{X}=\{\hat{\bm{x}}_1, \dots, \hat{\bm{x}}_{B-1}\}$ the already selected query points in this batch, where $B$ is the batch size. Note the corresponding observations ${\bm{y}}=\{{y}_1, \dots, {y}_{B-1}\}$ are unknown, since the simulations of this batch are not finished yet. Following the same penalization strategy as \cite{desautels2014parallelizing}, we assumes the underlying objective function follows the same behavior as the current predictive mean. Let $\hat{X}=\{\hat{\bm{x}}_1, \dots, \hat{\bm{x}}_{B-1}\}$ denote the query points under evaluations,  we approximate the observations with the predictive mean $\hat{\bm{y}}=\{\hat{y}_1, \dots, \hat{y}_{B-1}\}$ of GPR model with the observed data points $D=\{X, \bm{y}\}$ as training data. With the observed data points and the pseudo data points, i.e., $\hat D=\{X, \bm{y}\} \cup \{\hat{X}, \hat{\bm{y}}\}$, we could get the predictive uncertainty of GPR model $\hat \sigma(\bm{x})$ according to (\ref{eq:prediction}). And this uncertainty estimation $\hat \sigma(\bm{x})$ can naturally be incorporated into our acquisition function to guarantee the diversity of the query points in one batch. The acquisition function with penalization scheme can be expressed as:
\begin{equation}
    \alpha(\bm{x}, w) = (1-w)*\mu(\bm{x})+w*\hat \sigma(\bm{x}), 
\end{equation}
where $w=\kappa/(\kappa+1)$, and $\kappa$ is randomly sampled from the uniform distribution over a proper range $[0, \lambda]$. Here, the predictive uncertainty is replaced by the uncertainty estimation $\hat \sigma(\bm{x})$ to guarantee the exploration as well as diversity simultaneously.

\vspace{-0.1cm}
\subsection{Summary of EasyBO approach} \label{sec:summary}

We summarize EasyBO algorithm in Algorithm \ref{algo:summary}.

\begin{algorithm}[h]
    \caption{EasyBO Algorithm}
    \label{algo:summary}
    \begin{algorithmic}[1]
        \STATE Initialize a training dataset $D_o=\{X, \bm{y}\}$, and define the batch size $B$.
        \FOR{t=1 to N}
        \STATE Wait for a worker to be available.
        \STATE Update the training dataset with newly observed data $D_t=D_{t-1} \cup \{\bm{x}_t, y_t\}$.
        \STATE Get the predictive mean $\hat{\bm{y}}=\{\hat{y}_1, \dots, \hat{y}_{B-1}\}$ of the remaining samples $\hat{X}=\{\hat{\bm{x}}_1, \dots, \hat{\bm{x}}_{B-1}\}$ that are still under evaluation.
        \STATE Refine the surrogate model with $D_t \cup \{\hat{X}, \hat{\bm{y}}\}$.
        \STATE Propose the next candidate points $\hat{\bm{x}}_B$ for the idle worker by maximizing the acquisition function.
        \ENDFOR 
        \STATE Obtain the optimal results $\text{max}(\bm{y})$.
    \end{algorithmic}
\end{algorithm}

%% file: include/experiment.tex
\vspace{-0.1cm}
\section{Experimental Results} \label{sec:experiment}

In this section, we show the efficiency of EasyBO by comparing it with several state-of-the-art optimization algorithms in both sequential and batch modes, in terms of both the number of simulations and the wall-clock time \footnote{Since we only consider the cases when the objective function is expensive to evaluate, we exclude the time spent on modeling and selecting the candidate points.}. Our benchmark circuits include an operational amplifier (\S\ref{sec:opamp}) and a class-E power amplifier (\S\ref{sec:classE}), of which the simulation results are generated with the commercial HSPICE circuit simulator. All experiments are conducted on a Linux workstation with two Intel Xeon X5650 CPUs and 128GB memory.

Our sequential mode baselines include: (1) DE \cite{liu2009analog}, an optimization algorithm based on the evolutionary algorithm, (2) EI \cite{mockus1978application}, an improvement-based acquisition function that facilitate the optimization procedure of BO algorithm, (3) LCB \cite{srinivas2009gaussian}, an optimistic strategy that helps BO framework to fully explore the design space. The reason to evaluate EasyBO in sequential mode is to show that it still works reasonably well although it is not specifically designed for this mode.

For experiments in batch mode, we compare EasyBO against two state-of-the-art algorithms: pBO \cite{hu2018parallelizable} which combines both PI and LCB to select the candidate points on each batch, and its modified version pHCBO \cite{hu2018parallelizable} with high coverage consideration to penalize the clustered samples. To further investigate the relative merits of our proposed algorithm, we run another three alternatives derived from EasyBO: (1) EasyBO-S which selects the candidate points synchronously, (2) EasyBO-A which selects the candidate points asynchronously, (3) EasyBO-SP which synchronously selects the candidate points with our proposed penalization scheme to reduce sampling around the same region. EasyBO here represents our proposed asynchronous Bayesian optimization approach with our proposed penalization scheme. All the synchronous and asynchronous batch BO algorithms are presented in different batch sizes to demonstrate the robustness of our proposed algorithm. For simplicity of denotation, we label the algorithms with batch size in the tail. For pBO and pHCBO, we follow the same pattern of the weighting parameters $\bm{w}=(w_1, \dots, w_B)$ in \cite{hu2018parallelizable} and set $w_i = (i-1)/(B-1)$.

\vspace{-0.05cm}
\subsection{Operational Amplifier} \label{sec:opamp}

\begin{figure}
    \centering
    \includegraphics[width=0.35\textwidth]{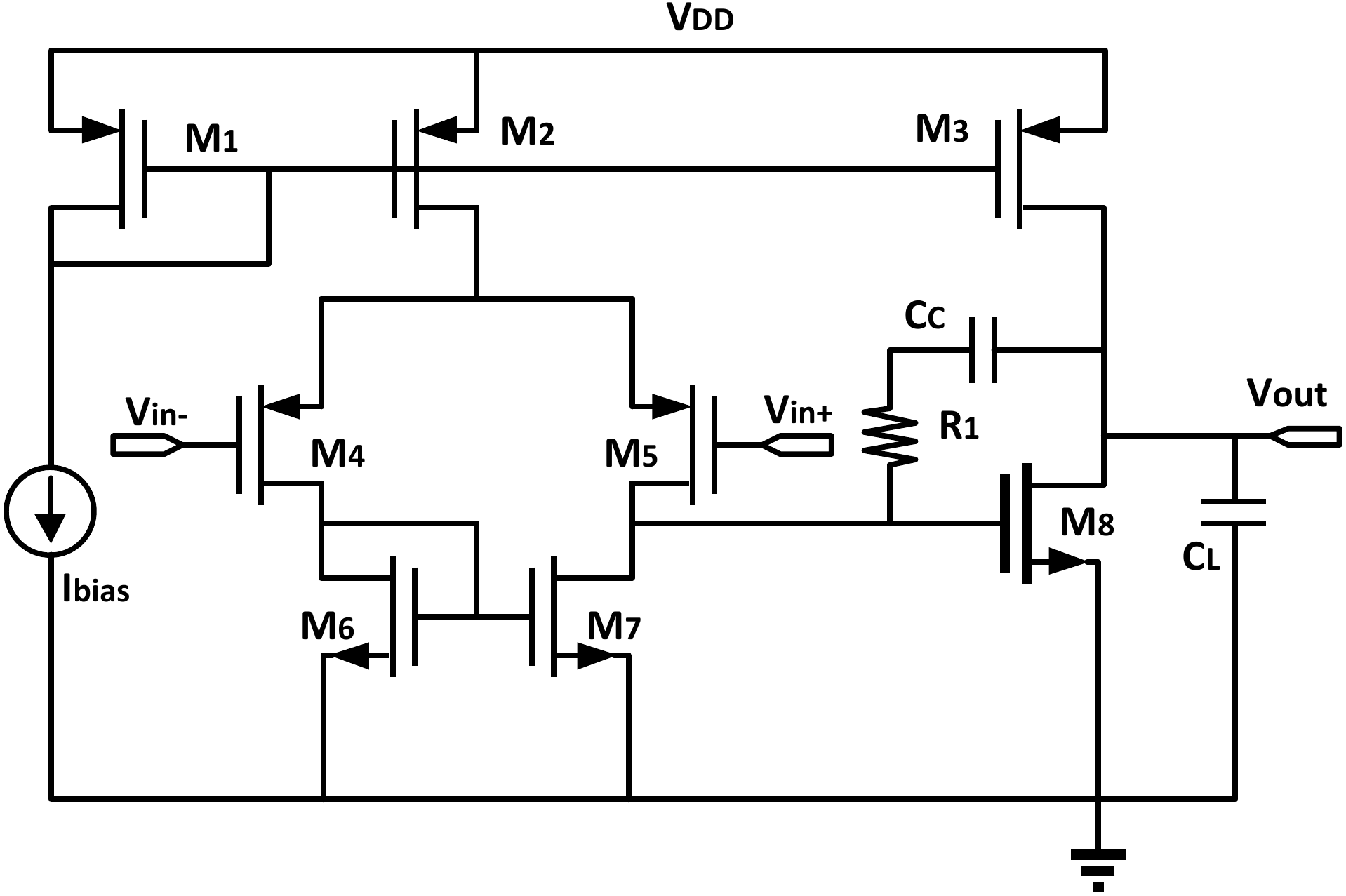}
    \vspace{-0.4cm}
    \caption{Schematic of the operational amplifier, which is reproduced from \cite{wang2014enabling}.}
    \label{fig:opamp_schematic}
    \vspace{-0.4cm}
\end{figure}

The operational amplifier circuit is implemented in a 180nm process, and its schematic is shown in Figure \ref{fig:opamp_schematic}. There is a total of 10 design variables, including the lengths and widths of the transistors, the resistance of the resistors, and the capacitance of the capacitors. The corresponding design specification is as follow:
\begin{equation}
    \text{maximize.} \quad 1.2 \times GAIN + 10 \times UGF + 1.6 \times PM, 
\end{equation}
where GAIN denotes the gain of the circuit, UGF means the unit gain frequency and PM is the phase margin.

To ensure a fair comparison, we run all the optimization algorithms 20 times to reduce the random fluctuations. For DE method, we set the maximum number of simulations to 20000. For algorithms that are based on the BO framework, we randomly sample 20 initial data points and limit the number of simulations to 150, regardless of the batch size. The optimization results and the corresponding time consumption on simulation are presented in Table \ref{tb:opamp}.  

\begin{table}
    \centering
    \caption{The optimization results and the corresponding simulation time of the operational amplifier circuit.}
    \label{tb:opamp}
    \begin{tabular}{cccccc}
        \hline
        Algo & Best & Worst & Mean & Std & Time \\
        \hline \hline
        DE & 685.44 & 680.654 & 682.19 & 1.56 & 216h40m51s \\
        LCB & 689.75 & 674.38 & 685.98 & 4.62 & 1h36m45s \\
        EI & 690.29 & 650.14 & 682.53 & 12.48 & 1h36m52s\\
        EasyBO & 690.36 & 688.10 & \textbf{689.87} & 0.92 & \textbf{1h36m55s} \\
        \hline \hline
        pBO-5 & 690.35 & 665.78 & 685.17 & 7.25 & 21m19s \\
        pHCBO-5 & 690.35 & 477.40 & 618.30 & 78.17 & 21m24s \\
        EasyBO-S-5 & 690.36 & 456.05 & 666.78 & 62.84 & 21m10s \\
        EasyBO-A-5 & 690.36 & 660.92 & 685.21 & 9.81 & 19m21s \\
        EasyBO-SP-5 & 690.36 & 688.10 & 689.97 & 0.78 & 21m6s \\
        EasyBO-5 & 690.36 & 688.10 & \textbf{690.17} & 0.58 & \textbf{19m10s} \\
        \hline \hline
        pBO-10 & 690.35 & 459.93 & 639.60 & 77.01 & 11m1s \\
        pHCBO-10 & 690.35 & 432.51 & 645.71 & 67.25 & 11m3s \\
        EasyBO-S-10 & 690.34 & 438.10 & 662.77 & 65.18 & 11m13s \\
        EasyBO-A-10 & 690.34 & 667.48 & 684.34 & 8.01 & 9m44s \\
        EasyBO-SP-10 & 690.36 & 673.34 & 688.73 & 4.34 & 11m4s \\
        EasyBO-10 & 690.36 & 688.10 & \textbf{689.84} & 0.79 & \textbf{9m42s} \\
        \hline \hline
        pBO-15 & 688.50 & 478.17 & 631.72 & 67.00 & 7m44s\\
        pHCBO-15 & 690.35 & 538.09 & 655.05 & 52.07 & 7m47s \\
        EasyBO-S-15 & 690.30 & 365.41 & 610.10 & 90.69 & 7m52s \\
        EasyBO-A-15 & 690.34 & 469.09 & 647.32 & 67.28 & 6m48s \\
        EasyBO-SP-15 & 690.36 & 673.21 & 687.78 & 4.77 & 7m47s \\
        EasyBO-15 & 690.36 & 678.98 & \textbf{688.09} & 3.18 & \textbf{6m43s} \\
        \hline
    \end{tabular}
    \vspace{-0.6cm}
\end{table}

We start with examining the performance of EasyBO in sequential mode. Compared with EI and LCB, EasyBO achieves better optimization results while spending similar time on simulation, as is shown in the top block of Table \ref{tb:opamp}. Compared with DE, EasyBO reduces the simulation time by 134$\times$ while obtaining better optimization results. This clearly shows the EasyBO is efficient and effective even in the non-batch mode.

Now let's move on to the batch mode results to witness the full potential of EasyBO. Results under 3 representative batch sizes (5, 10 and 15) are presented to understand the impact of different batch size. Overall, EasyBO constantly achieves better optimization results and less simulation time with respect to the same batch size. In other words, EasyBO has higher sample efficiency in terms of both the number of simulations and wall-clock time for the same batch size. Considering the impact of asynchronous sampling, asynchronous batch BO algorithm is able to process a greater number of simulations in a given period of time, intuitively. The experimental results further give an empirical demonstration that asynchronous batch BO algorithm has a higher hardware utilization compared to its synchronous counterpart. Also, the alternatives with our proposed penalization scheme (EasyBO and EasyBO-SP) constantly outperforms EasyBO-S and EasyBO-A, which demonstrates that our proposed penalization scheme helps to prevent redundant samples from being chosen and significantly improve the sample efficiency.

To further analyze the impact of the batch size, we fix the algorithm and compare the optimization results across different batch size. For pBO, pHCBO, EasyBO-S and EasyBO-A, the optimization results tend to deteriorate rapidly with the increase of the batch size, which means that the sample efficiency decreases quickly as the batch size increases. Instead, the optimization results of both EasyBO and EasyBO-SP are much more stable across different batch size, which means that our proposed penalization scheme helps to maintain the sample efficiency as the batch size increases. An interesting yet unexpected behavior we note is that the optimization results of EasyBO in batch mode is even competitive compared to its sequential counterpart. This observation further demonstrates that a relative magnitude of batch size encourages exploration. Also, from a time perspective, EasyBO generally reduces $B\times$ of the simulation time compared to the sequential EasyBO.

\begin{figure}
    \centering
    \includegraphics[width=0.34\textwidth]{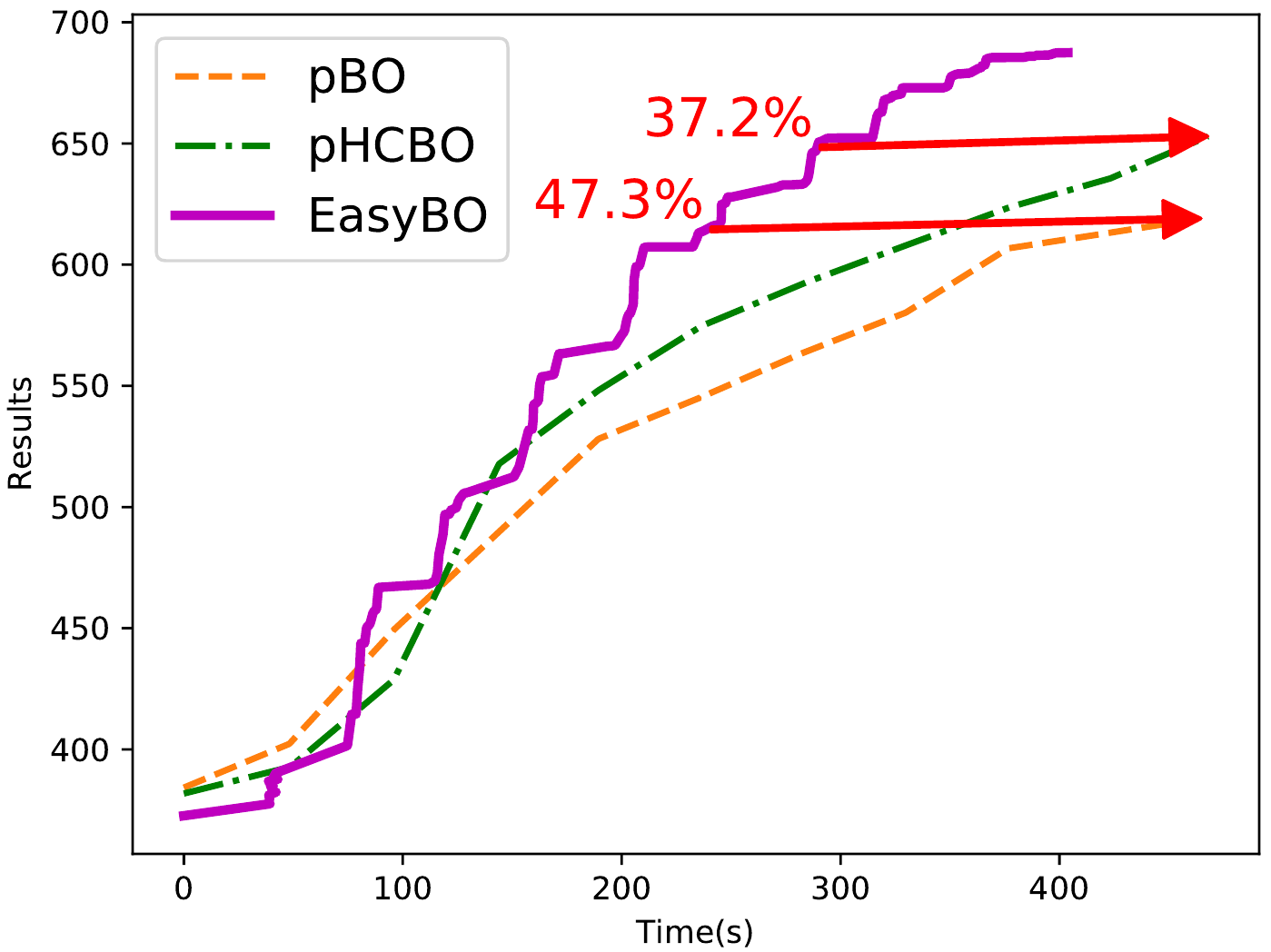}
    \vspace{-0.4cm}
    \caption{The optimization results of the operational amplifier circuits v.s. the wall-clock time, when the batch size is 15.}
    \label{fig:opamp_time}
    \vspace{-0.3cm}
\end{figure}

The above results empirically demonstrate the efficiency and effectiveness of our proposed EasyBO algorithm. Without sacrificing the optimization results, EasyBO can achieve up to 1935$\times$ speed-up compared with DE algorithm for B=15. For a fixed number of simulations, 9.2\%, 12.7\% and 13.7\% of the time reduction can be achieved by EasyBO compared with its synchronous counterpart, when the batch size is 5, 10 and 15 respectively. The time reduction increases as the batch size increases, due to the increased idle time of the synchronous batch BO framework. In other words, the time reduction of the synchronous batch BO algorithms efficiency decreases as the batch size increases. In Figure \ref{fig:opamp_time}, we present the optimization results in terms of wall-clock time for B=15. With the same optimization result, our proposed EasyBO can reduce 47.3\% and 37.4\% of the simulation time, compared with pBO and pHCBO respectively. Experimental results from above demonstrate that EasyBO has a higher sample efficiency in terms of both the number of simulations and wall-clock time.


\vspace{-0.05cm}
\subsection{Class-E Power Amplifier} \label{sec:classE}

\begin{figure}
\centering
\includegraphics[width=0.34\textwidth]{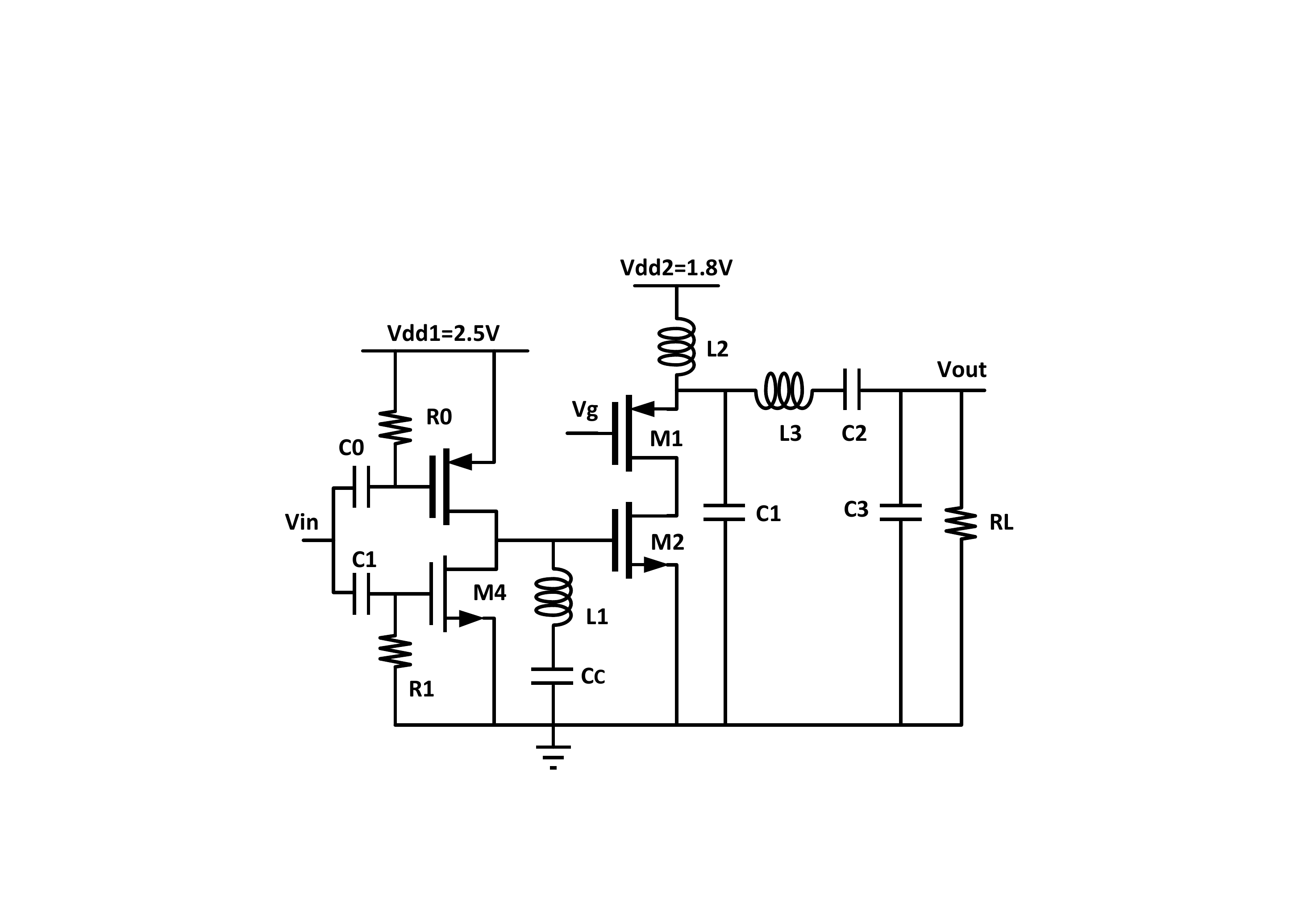}
\vspace{-0.4cm}
\caption{Schematic of the class-E power amplifier, which is reproduced from \cite{lyu2018batch}.}
\label{fig:classE_schematic}
\vspace{-0.3cm}
\end{figure}

Implemented in a 180nm process, the schematic of the class-E power amplifier is presented in Figure \ref{fig:classE_schematic}. There are 12 design parameters in total, and the corresponding design specification is constructed as:
\begin{equation}
    \text{maximize.} \quad 3 \times PAE + Pout,
\end{equation}
where PAE is the power added efficiency and Pout means the output power.

Again, to ensure a fair comparison, we run each optimization algorithm 20 times to average the random fluctuations. The maximum number of simulations for DE methodology is set to 15000. For the remaining algorithms, a total of 20 initial data points are randomly sampled and the maximum number of simulations is set to 450. The experimental results and the corresponding simulation time of the class-E power amplifier circuit are presented in Table \ref{tb:classE}.

\begin{table}
    \centering
    \caption{The optimization results and the corresponding simulation time of the class-E circuit.}
    \label{tb:classE}
    \begin{tabular}{cccccc}
        \hline
        Algo & Best & Worst & Mean & Std & Time \\
        \hline \hline
        DE & 4.56 & 4.33 & 4.43 & 0.08 & 220h20m40s \\
        LCB & 4.10 & 3.59 & 3.89 & 0.14 & 6h35m41s \\
        EI & 4.13 & 3.52 & 3.85 & 0.19 & 6h36m3s \\
        EasyBO & 5.15 & 4.12 & \textbf{4.44} & 0.24 & \textbf{6h35m18s} \\
        \hline \hline
        pBO-5 & 4.61 & 3.76 & 4.17 & 0.19 & 1h48m24s \\
        pHCBO-5 & 4.42 & 3.66 & 4.17 & 0.16 & 1h48m32s \\
        EasyBO-S-5 & 5.36 & 4.04 & \textbf{4.54} & 0.39 & 2h0m17s \\
        EasyBO-A-5 & 5.25 & 3.56 & 4.32 & 0.37 &  1h19m44s \\
        EasyBO-SP-5 & 5.08 & 3.96 & 4.47 & 0.25 & 1h49m11s \\
        EasyBO-5 & 4.62 & 4.10 & 4.40 & 0.18 & \textbf{1h19m26s} \\
        \hline \hline
        pBO-10 & 4.34 & 3.80 & 4.11 & 0.16 & 1h2m49s \\
        pHCBO-10 & 4.82 & 3.79 & 4.17 & 0.23 & 1h2m51s \\
        EasyBO-S-10 & 5.51 & 3.54 & 4.23 & 0.48 & 58m11s \\
        EasyBO-A-10 & 4.78 & 3.92 & 4.26 & 0.19 & 40m41s \\
        EasyBO-SP-10 & 5.12 & 4.15 & 4.52 & 0.24 & 55m6s \\
        EasyBO-10 & 5.12 & 4.15 & \textbf{4.53} & 0.28 & \textbf{40m16s} \\
        \hline \hline
        pBO-15 & 4.61 & 3.87 & 4.17 & 0.19 & 44m6s \\
        pHCBO-15 & 4.31 & 3.67 & 4.10 & 0.16 & 51m49s \\
        EasyBO-S-15 & 5.14 & 3.17 & 4.14 & 0.43 & 41m43s \\
        EasyBO-A-15 & 5.41 & 3.59 & 4.34 & 0.43 & 27m14s \\
        EasyBO-SP-15 & 5.08 & 4.20 & 4.47 & 0.25 & 41m43s \\
        EasyBO-15 & 5.74 & 4.22 & \textbf{4.51} & 0.32 & \textbf{26m24s}\\
        \hline
    \end{tabular}
    \vspace{-0.4cm}
\end{table}

\begin{figure}
    \centering
    \includegraphics[width=0.35\textwidth]{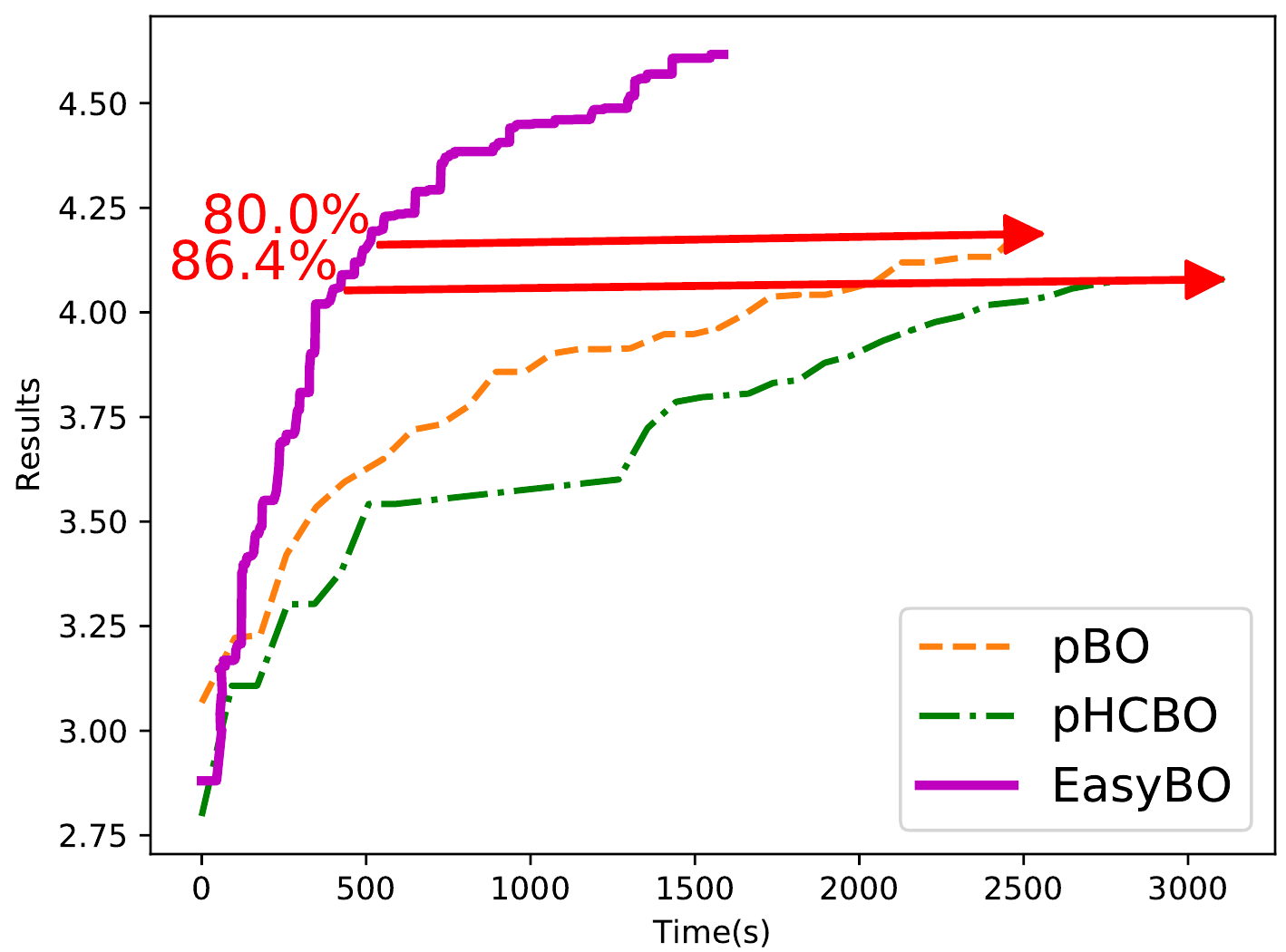}
    \vspace{-0.3cm}
    \caption{The optimization results of the class-E power amplifier circuit v.s. the wall-clock time, when the batch size is 15.}
    \label{fig:classE_time}
    \vspace{-0.3cm}
\end{figure}

We again start with examining the performance of EasyBO in the sequential mode. Compared to EI and LCB, EasyBO achieves better optimization results while spending similar time on simulation. Compared to DE, EasyBO reduces up to 33$\times$ of the simulation time with better optimization results. 

In the batch mode, EasyBO and its two alternatives EasyBO-A and EasyBO-SP constantly achieve better optimization results compared with both pBO and pHCBO. An interesting observation is that the optimization results of both EasyBO and EasyBO-SP when batch size is 10 and 15 are better than when batch size is 5 and 1. This means our proposed penalization scheme encourages exploration with relative batch size. From a wall-clock time perspective, EasyBO reduces the time consumption on simulation by up to 500$\times$ compared with DE methodology, while obtaining better optimization results. Also, with a fixed number of simulations, EasyBO reduces approximately 26.7\%, 35.7\% and 40.0\% of the simulation time compared with pBO and pHCBO when batch size is 5, 10 and 15 respectively. As is shown in Figure \ref{fig:classE_time}, with the same optimization result, EasyBO can reduce 80.0\% and 86.4\% of the simulation time for B=15, compared with pBO and pHCBO respectively. In other words, EasyBO can achieve up to 7.35$\times$ speed-up, with comparable optimization results. Experimental results demonstrate that EasyBO is more efficient and effective in terms of both the number of simulations and wall-clock time.

%% file: include/conclusion.tex
\vspace{-0.1cm}
\section{conclusion} \label{sec:conclusion}

EasyBO is a novel asynchronous batch Bayesian optimization approach for analog circuit synthesis. By asynchronously proposing the next query point, EasyBO makes a higher utilization of the hardware resources. We developed a new acquisition function which can better explore the design space for asynchronous batch Bayesian optimization. And we further penalize around the sampled area to enhance the sample efficiency. Despite being an efficient asynchronous batch BO algorithm, EasyBO also provides backward compatibility to work in both sequential mode and synchronous batch mode. Compared to the state-of-the-art synchronous batch BO algorithms, EasyBO achieves up to 7.35$\times$ speed-up with comparable optimization results.